\documentclass{aastex631}

\newif\ifarxiv

\begin{document}
\arxivtrue

\title{Automatic detection of plateau phases in light curves of variable stars}

\author{Anastasia Lavrukhina}
\affiliation{Faculty of Space Research, Lomonosov Moscow State University,
\\ Leninsky Gori 1 bld. 52, Moscow 119234, Russia}
\author[0000-0001-7179-7406]{Konstantin Malanchev}
\affiliation{Department of Astronomy, University of Illinois at Urbana-Champaign, \\ 1002 W. Green St., IL 61801, USA}
\affiliation{Sternberg Astronomical Institute, Lomonosov Moscow State University, \\ Universitetsky 13, Moscow 119234, Russia}
\affiliation{ McWilliams Center for Cosmology, Department of Physics, Carnegie Mellon University, \\ Pittsburgh, PA 15213, USA}
\author{Matwey V. Kornilov}
\affiliation{Sternberg Astronomical Institute, Lomonosov Moscow State University, \\ Universitetsky 13, Moscow 119234, Russia}
\affiliation{National Research University Higher School of Economics, \\ 21/4 Staraya Basmannaya Ulitsa, Moscow, 105066, Russia}



\begin{abstract}

Modern astronomical surveys produce millions of light curves of variable sources.
These massive data sets challenge the community to create automatic light-curve processing methods for detection, classification, and characterisation of variable stars.
In this paper, we present a novel method for extracting the variable components of a light curve based on Otsu's thresholding method.
To validate the effectiveness of this method, we apply it to the light curves of detached eclipsing binaries and dwarf novae, sourced from OGLE catalogues.

\end{abstract}



\section{Introduction} \label{sec:intro}

Astronomy has entered an era marked by large surveys of variable sources, such as the Zwicky Transient Facility~\citep{Belm2019} and the upcoming Vera C. Rubin Observatory Legacy Survey of Space and Time~\citep{LSST}.
These surveys aim to observe variable stellar objects with a cadence of a few days, generating up to a million alerts nightly.
Handling such volumes of data demands automated and efficient processing techniques to pave the way for novel discoveries.
One of the common approaches is an extraction of a number of light-curve features with their subsequent use for objects classification and characterization.
Today, numerous feature extractors can be accessed through public programming libraries like \texttt{Feets}\citep{feets} or the Alerce classifier feature extractor\citep{ALERCE}.
Nevertheless, devising new features tailored for detecting specific types of objects remains a significant challenge.

Within our library, \texttt{light-curve}\footnote{https://github.com/light-curve} \citep{lightcurve, 2021MNRAS.502.5147M}, we have explored several features that best characterize the light curves of variable stars.
Many features employed in comparable research aim to describe the magnitude (or flux) distribution of observations, such as the $n^\mathrm{th}$ moment, interquartile range, and others.
They also attempt to outline the shape of light curves~\citep{Kim, Stetson_1996}.
However, these features often overlook the asymmetry of the magnitude distribution.
In other words, they may not effectively differentiate between sources that are becoming brighter (out-bursting) and those that are dimming.
For instance, low-amplitude dwarf novae have values of "symmetric" light-curve features similar to eclipsing binaries.
Yet, the latter constitute a substantial portion of all known variable stars~\citep{vsx}; this similarity can lead to misclassifications, especially when the periodicity of the light curve cannot be detected.
Hence, there's a pressing need for features that can delineate the disparities between the variable and constant segments of a light curve.

In this study, we explore possible applications of Otsu's method~\citep{Otsu}, originally developed for image processing and computer vision to address object detection challenges.
The idea of this method is the determination of an optimal brightness threshold that separates the object from its background.
We suggest applying this method to a sample of light-curve magnitudes, with the aim to isolate the variable portion of the light curve from its constant counterpart.
In this paper, we present the first time application of Otsu's method to variable star light curves.
We showcase the results of its efficacy when applied to dwarf novae and eclipsing binaries as observed by the Optical Gravitational Lensing Experiment (OGLE) \citep{ogle3, ogle4}.

\section{Otsu's method} \label{sec:style}

Otsu's method is used in the field of computer vision for the purpose of image binarization.
This involves dividing the original sample of pixel brightness using a threshold specifically chosen to minimize the intra-class variance, denoted as $\sigma^2_{W}$\ref{lavrukhina:eq1}. Otsu demonstrated that this approach is analogous to maximizing the inter-class variance, represented by $\sigma^2_{B}$\ref{lavrukhina:eq2}.

\begin{equation}\label{lavrukhina:eq1}
\sigma^2_{W}=w_0\sigma^2_0+w_1\sigma^2_1 \,,
\end{equation}
\begin{equation}\label{lavrukhina:eq2}
\sigma^2_{B}=w_0 w_1 (\mu_1-\mu_0)^2 \,,
\end{equation}
where $w_i$ is the weight of the sample (ratio of the subsample size to the overall quantity of objects); $\sigma^2_i$ is the dispersion of the subsample; $\mu_i$ is the subsample mean; $i = {0, 1}$ denotes "bright" and "faint" subsamples (i.e. having lower and larger magnitude values).
We assume that the maximum light of an out-bursting object would be in the bright subsample while the minimum light of a dimming object would be in the faint subsample.

\section{Data} \label{sec:floats}

We use the OGLE $I$-passband light curves for both dwarf novae and eclipsing binaries.
The light curves of dwarf novae are taken from the OGLE catalogue "One Thousand New Dwarf Novae"~\citep{mroz2016thousand}. 
For the eclipsing binaries, we refer to the OGLE-III Galactic Disk Fields\citep{pietrukowicz2013eclipsing}. 
We selected detached eclipsing binaries only ("nonEC" type and period larger than one day) ensuring we capture sharp eclipse profiles.

\section{Results} \label{subsec:tables}

\begin{figure}[ht!]
\plotone{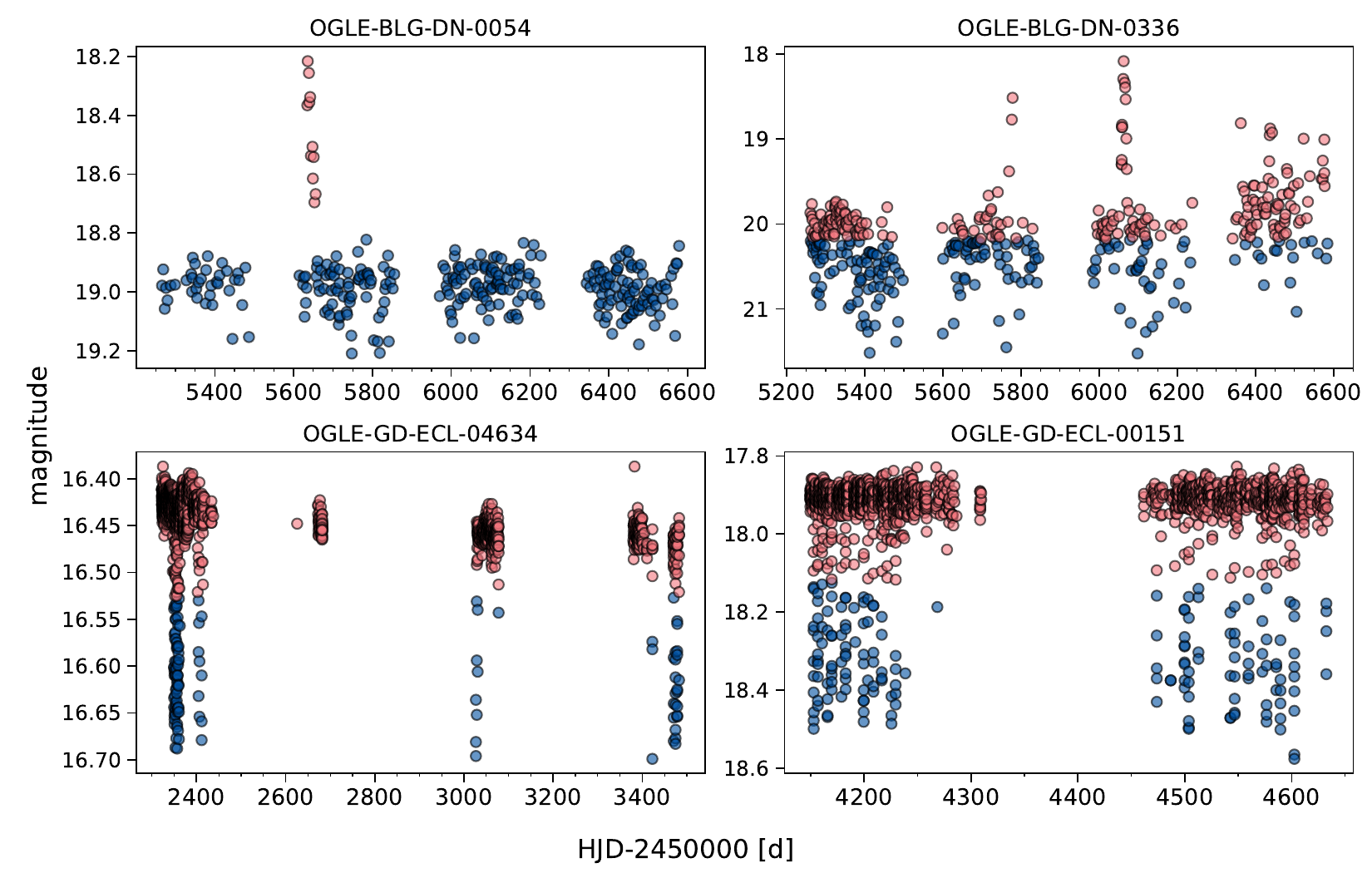}
\caption{Results of the method applied to light curves of dwarf novae (upper panels) and eclipsing binaries (lower panels): blue colour shows faint subset, red colour shows bright subset.
\label{fig:otsu_example}}
\end{figure}

Fig.~\ref{fig:otsu_example} shows examples of thresholds determined using Otsu's method.
The left panels illustrate successful splits between the light curve's maximum and minimum, especially evident when outbursts (or eclipses) exhibit a large amplitude and a notable total duration.
Conversely, the right panels indicate that when substantial low-amplitude variability is present, the appearance of a plateau becomes less discernible, leading to suboptimal performance by the method.

\ifarxiv
    In Fig.~\ref{fig:otsu_distr}, we present the distributions of features extracted from both the faint and bright subsamples of the light curves.
    The top left panel shows the difference between mean magnitudes of the subsamples $(\mu_0 - \mu_1)$.
    This difference in distribution arises primarily from the inclusion of high-amplitude dwarf novae in the dataset.
    The top right panel presents the ratio of observations in the bright subsample to the entire observation count.
    It's evident that there's a higher concentration of observations in the subsample expected to contain a plateau.
    Consequently, for dwarf novae, the faint subsamples are more predominant, while for eclipsing binaries, the bright subsamples are more populated.
    The lower panels display the distributions of the standard deviation for both bright and faint subsamples.
    Eclipsing binaries show a narrower distribution for the bright subsample, which can be explained by a smaller variance of the subsample, containing a plateau.

    \begin{figure}[ht!]
    \plotone{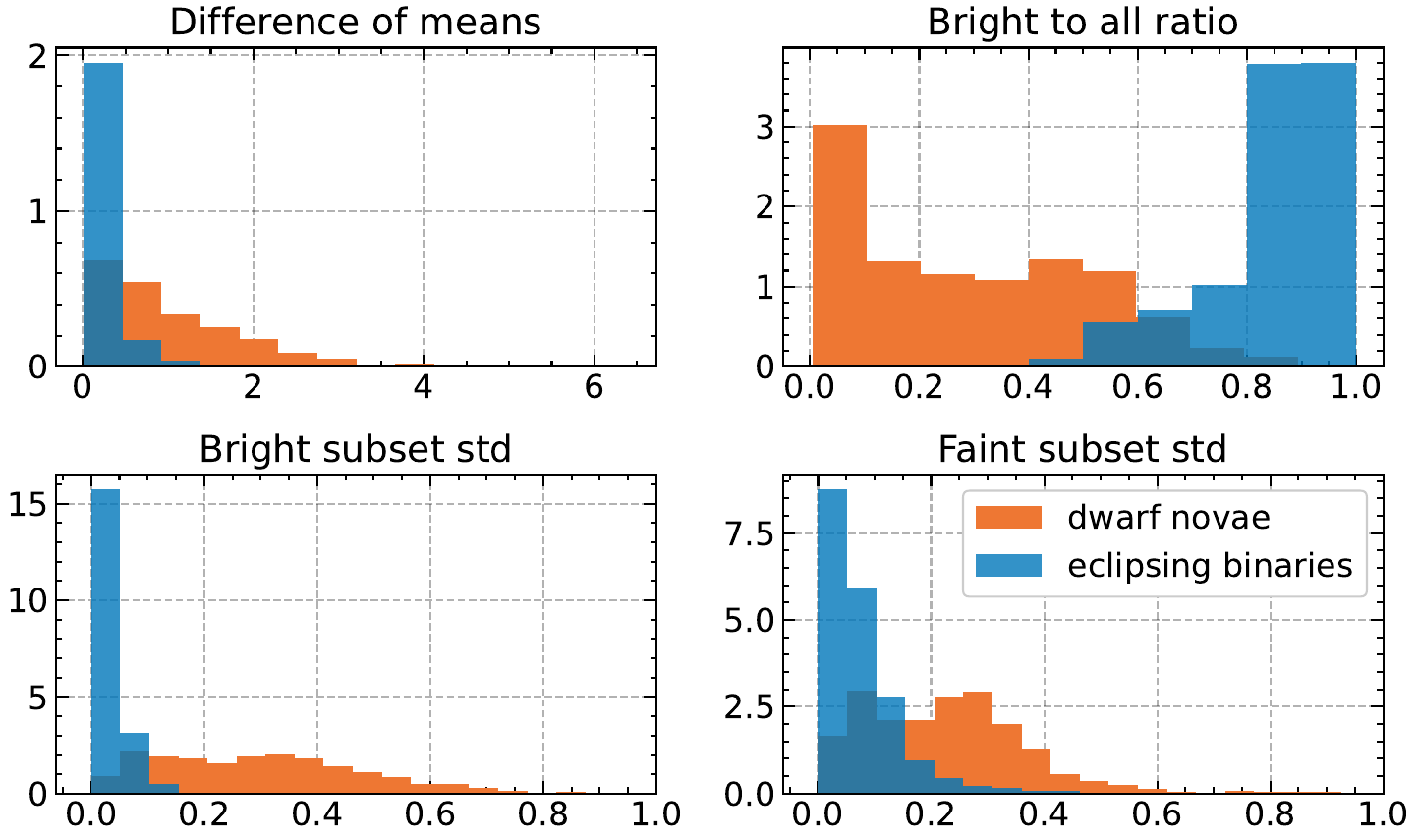}
    \caption{Histograms of feature's distributions, extracted from the faint and bright subsamples: the difference between the mean magnitude of bright and faint subsample, the ratio of number of observations in faint subsample to the total number of observations, standard deviations in bright and faint subsamples.
    \label{fig:otsu_distr}}
    \end{figure}
\fi

In summary, our research introduces the innovative application of Otsu's method to the field of variable star photometric classification.
We reveal that the newly proposed features provide a nuanced understanding of light curve characteristics, effectively distinguishing between different types of variable stars.
The incorporation of these features into existing classification and characterization algorithms holds promise for significantly enhancing their accuracy and reliability.

\begin{acknowledgments}
The reported study was funded by
RFBR and CNRS according to the research project №21-52-15024.
\end{acknowledgments}

\bibliography{lavrukhina}{}

\begin{thebibliography}{}
\expandafter\ifx\csname natexlab\endcsname\relax\def\natexlab#1{#1}\fi
\providecommand{\url}[1]{\href{#1}{#1}}
\providecommand{\dodoi}[1]{doi:~\href{http://doi.org/#1}{\nolinkurl{#1}}}
\providecommand{\doeprint}[1]{\href{http://ascl.net/#1}{\nolinkurl{http://ascl.net/#1}}}
\providecommand{\doarXiv}[1]{\href{https://arxiv.org/abs/#1}{\nolinkurl{https://arxiv.org/abs/#1}}}

\bibitem[{{Bellm} {et~al.}(2019)}]{Belm2019}
{Bellm}, E.~C., {et~al.} 2019, \pasp, 131, 018002,
  \dodoi{10.1088/1538-3873/aaecbe}

\bibitem[{{Cabral} {et~al.}(2018){Cabral}, {Sanchez}, {Ramos}, {Gurovich},
  {Granitto}, \& {VanderPlas}}]{feets}
{Cabral}, J., {Sanchez}, B., {Ramos}, F., {et~al.} 2018, {feets: feATURE
  eXTRACTOR for tIME sERIES}.
\newblock \doeprint{1806.001}

\bibitem[{{Kim, Dae-Won} {et~al.}(2014){Kim, Dae-Won}, {Protopapas, Pavlos},
  {Bailer-Jones, Coryn A. L.}, {Byun, Yong-Ik}, {Chang, Seo-Won}, {Marquette,
  Jean-Baptiste}, \& {Shin, Min-Su}}]{Kim}
{Kim, Dae-Won}, {Protopapas, Pavlos}, {Bailer-Jones, Coryn A. L.}, {et~al.}
  2014, A\&A, 566, A43, \dodoi{10.1051/0004-6361/201323252}

\bibitem[{{Lavrukhina} \& {Malanchev}(2021)}]{lightcurve}
{Lavrukhina}, A.~D., \& {Malanchev}, K.~L. 2021, in Proceedings of Astronomy
  and space exploration 2021, UrFU (Ekaterinburg: Ural University Press),
  133--136, \dodoi{10.15826/B978-5-7996-3229-8.32}

\bibitem[{{LSST Science Collaboration} {et~al.}(2009)}]{LSST}
{LSST Science Collaboration}, {et~al.} 2009, arXiv e-prints, arXiv:0912.0201.
\newblock \doarXiv{0912.0201}

\bibitem[{{Malanchev} {et~al.}(2021){Malanchev}, {Pruzhinskaya}, {Korolev},
  {Aleo}, {Kornilov}, {Ishida}, {Krushinsky}, {Mondon}, {Sreejith}, {Volnova},
  {Belinski}, {Dodin}, {Tatarnikov}, {Zheltoukhov}, \& {(The SNAD
  Team)}}]{2021MNRAS.502.5147M}
{Malanchev}, K.~L., {Pruzhinskaya}, M.~V., {Korolev}, V.~S., {et~al.} 2021,
  \mnras, 502, 5147, \dodoi{10.1093/mnras/stab316}

\bibitem[{Mroz {et~al.}(2016)Mroz, Udalski, Poleski, Pietrukowicz, Szymanski,
  Soszynski, Wyrzykowski, Ulaczyk, Kozlowski, \& Skowron}]{mroz2016thousand}
Mroz, P., Udalski, A., Poleski, R., {et~al.} 2016, One Thousand New Dwarf Novae
  from the OGLE Survey.
\newblock \doarXiv{1601.02617}

\bibitem[{Otsu(1979)}]{Otsu}
Otsu, N. 1979, IEEE Transactions on Systems, Man, and Cybernetics, 9, 62,
  \dodoi{10.1109/TSMC.1979.4310076}

\bibitem[{Pietrukowicz {et~al.}(2013)Pietrukowicz, Mroz, Soszynski, Udalski,
  Poleski, Szymanski, Kubiak, Pietrzynski, Wyrzykowski, Ulaczyk, Kozlowski, \&
  Skowron}]{pietrukowicz2013eclipsing}
Pietrukowicz, P., Mroz, P., Soszynski, I., {et~al.} 2013, Eclipsing Binary
  Stars in the OGLE-III Galactic Disk Fields.
\newblock \doarXiv{1306.6324}

\bibitem[{Stetson(1996)}]{Stetson_1996}
Stetson, P.~B. 1996, Publications of the Astronomical Society of the Pacific,
  108, 851, \dodoi{10.1086/133808}

\bibitem[{Sánchez-Sáez {et~al.}(2021)}]{ALERCE}
Sánchez-Sáez, P., {et~al.} 2021, The Astronomical Journal, 161, 141,
  \dodoi{10.3847/1538-3881/abd5c1}

\bibitem[{{Udalski} {et~al.}(2008){Udalski}, {Szymanski}, {Soszynski}, \&
  {Poleski}}]{ogle3}
{Udalski}, A., {Szymanski}, M.~K., {Soszynski}, I., \& {Poleski}, R. 2008, Acta
  Astronomica, 58, 69.
\newblock \doarXiv{0807.3884}

\bibitem[{{Udalski} {et~al.}(2015){Udalski}, {Szyma{\'n}ski}, \&
  {Szyma{\'n}ski}}]{ogle4}
{Udalski}, A., {Szyma{\'n}ski}, M.~K., \& {Szyma{\'n}ski}, G. 2015, Acta
  Astronomica, 65, 1.
\newblock \doarXiv{1504.05966}

\bibitem[{{Watson} {et~al.}(2006){Watson}, {Henden}, \& {Price}}]{vsx}
{Watson}, C.~L., {Henden}, A.~A., \& {Price}, A. 2006, Society for Astronomical
  Sciences Annual Symposium, 25, 47

\end{thebibliography}
\bibliographystyle{aasjournal}



\end{document}